\documentclass[preprint,showpacs,amsmath,amssymb]{revtex4}
\usepackage{amsmath,amssymb,latexsym,graphics,epsfig}
\begin{document}

\title{Extreme statistics for time series: \\
Distribution of the maximum relative to the initial value}

\author{T. W. Burkhardt}
\email{tburk@temple.edu}
\affiliation{Department of Physics, Temple University, Philadelphia 19122, USA}

\author{G. Gy\"{o}rgyi}
\email{gyorgyi@glu.elte.hu}
\author{N. R. Moloney}
\email{moloney@general.elte.hu}
\author{Z. R\'{a}cz}
\email{racz@general.elte.hu}

\affiliation{Institute for Theoretical Physics - HAS,
  E\"{o}tv\"{o}s University, P\'{a}zm\'{a}ny
  s\'{e}t\'{a}ny 1/a, 1117 Budapest, Hungary}
\date{\today}

\begin{abstract}
The extreme statistics of time signals is studied when the maximum is
measured from
the initial value. In the case of independent,
identically distributed (iid)
variables, we classify the limiting distribution of the maximum according
to the properties of the parent distribution from which the variables
are drawn. Then we
turn to correlated periodic Gaussian signals with a $1/f^\alpha$ power
spectrum and
study the distribution of the maximum relative height with respect
to the initial
height (MRH$_{\rm I}$). The exact MRH$_{\rm I}$ distribution is derived
for $\alpha=0$ (iid
variables), $\alpha=2$ (random walk), $\alpha=4$ (random acceleration),
and $\alpha=\infty$ (single sinusoidal mode). For other, intermediate values
of $\alpha$, the distribution is determined from simulations. We find that
the MRH$_{\rm I}$ distribution is markedly different from the previously
studied distribution of
the maximum height relative to the average height for all $\alpha$.
The two main distinguishing features of the MRH$_{\rm I}$ distribution
are the much larger weight
for small relative heights and the divergence at zero height for
$\alpha>3$. We also demonstrate that the boundary conditions affect
the shape of the distribution by
presenting exact results for some non-periodic boundary conditions.
Finally, we show
that, for signals arising from time-translationally invariant distributions,
the
density of near extreme states is the same as the MRH$_{\rm I}$
distribution. This
is used in developing a scaling theory for the threshold singularities
of the two distributions.
\end{abstract} 

\maketitle

\section{Introduction}
The importance of extreme value statistics \cite{ft,gn,gumbel,galambos}
has long been recognized in engineering fields such as
hydrology \cite{hydro-2002}, as well as in insurance and
finance \cite{EVS-finance},
where gauging the effects of catastrophic events is a central concern.
Fascination with catastrophic events has also brought extreme statistics
into the
focus of everyday interest, as witnessed by debates about climatic events,
such as the most violent tornado or the hottest summer of the last
century \cite{Solow-1989}. In physics, on
the other hand, the use of extreme value statistics has not been widespread.
The reason for this may be that the rarity of the extreme events
naturally puts a high price on obtaining information. Nevertheless,
the last decade has seen an increasing
interest in extreme value statistics in physical applications, related, for example, to the ground state of spin glasses \cite{Mezard-97},
to interface fluctuations
\cite{Raych:2001,Antal:2001,Gyorgyi:2003,mc,Bolech:2004,Guclu:2004},
to fragmentation problems \cite{Krapivsky:2000},
to level-density problems of ideal quantum gases \cite{Comtet:2007},
to atmospheric physics \cite{Bunde-2006}, etc.

Some of these applications involve extensions of mathematically well known
results for independent, identically distributed (iid) variables
\cite{ft,gn,gumbel,galambos} to the physically relevant case of strongly
correlated variables. This has lead to some exactly solved examples and
simulation studies of particular systems with well known
correlations. A systematic classification
of the effect of correlations has, however, not yet emerged.

In the simplest case, extreme value statistics
emerges from random numbers drawn from a given distribution without
any reference to the order of the draws. Often, however, the
extremum is selected from a well ordered set of random variables.
For example, the extremum may be the maximum of a time series
$h(t)$ in the interval $0<t<T$ or, equivalently, the maximum height of
an interface $h(t)$ in the two-dimensional space $(t,h)$. A relevant
point to note here is that in correlated systems the boundary conditions at
$t=0$ and $t=T$ may be important. In particular, it has been
demonstrated for one-dimensional interfaces with periodic or free
boundary conditions that, in the presence of strong
correlations, the boundary conditions do affect the extreme value
distribution \cite{mc}.

The sensitivity to boundary conditions brings up the question whether
the extreme statistics also depends on the zero level from which the
maximum is measured. In the iid case, the maximum is usually
specified with respect to a fixed zero level, related to the scale of
parent distribution. In other cases, however, one may want
to define the zero level through some average measured in the random system.
For example, in the case of an interface, it is convenient to specify
the maximum height of a given realization $h(t)$ with respect to its
average $\overline h=T^{-1}\int_0^T h(t)dt$, which is
also a fluctuating variable \cite{Raycha:2001}.
In cases where the fluctuations of $\bar h$
diverge in the limit $T\to\infty$, it is not surprising that
the extreme height distribution is
sensitive to the choice of the zero level.

In physical applications, choosing the origin at the first value of the measurement,
i.e., measuring the maximum height with respect to the initial height,
is one of
several natural possibilities. In finance, as well, one may be interested
in the probable extremes of stock prices with respect to the starting price. Measuring the height of
a signal from the initial height instead of the average height, for example,
may seem like a trivial shift of the origin. However, if the initial
height is a random
variable, with a distribution determined either by the experimental
setup or by the
inherent statistical properties of the infinite signal,
then the extreme statistics
may depend on the probability distribution of the initial value.

In this paper we study the distribution of the quantity $m={\rm max}_t[h(t)-h(0)]$,
i.e., of the maximum height relative to the initial height (MRH$_{\rm I}$).
After a
comprehensive review of the iid case, we investigate $m$ for correlated Gaussian
signals $h(t)$ with a $1/f^\alpha$ power spectrum and with periodicity
$h(t)=h(t+T)$. It is interesting to compare the MRH$_{\rm I}$ distribution with the
distribution of the quantity $h_m={\rm max}_t[h(t)-\overline h]$, i.e., of the
maximum height relative to the average height (MRH$_{\rm A}$), recently analyzed in
Refs. \cite{mc,gyorgyietal}. One of our main findings is that the MRH$_{\rm I}$ and
MRH$_{\rm A}$ distributions are different for all $\alpha$.

The paper is organized as follows: The case of independent, identically distributed
(iid) variables is considered in Sec.~II. We show that the MRH$_{\rm I}$
distribution is the same as the recently studied density of near extreme events
\cite{sm}, and we find that the MRH$_{\rm I}$ and MRH$_{\rm A}$ distributions may or
may not differ, depending on the tail of the parent distribution. The model of
periodic, correlated, Gaussian signals $h(t)$ with a $1/f^\alpha$ noise spectrum is
introduced in Sec.~III, and our procedure for determining the MRH$_{\rm I}$
distribution from simulations is described. In Sec.~IV the exact MRH$_{\rm I}$
distribution is derived in the special cases $\alpha=0$ (iid Gaussian variables),
$\alpha=2$ (random walk), $\alpha=4$ (random acceleration), and $\alpha=\infty$
(single sinusoidal mode), including the dependence of the distribution
on the boundary conditions for $\alpha=2$ and $\alpha=4$.
Simulation results for the MRH$_{\rm I}$ distribution with periodic boundary
conditions for
other, intermediate values of $\alpha$ are presented in Sec.~V, and the
evolution of
the distribution with changing $\alpha$ is discussed. In Sec. VI we show that the
equivalence of the density of near extreme events and the MRH$_{\rm I}$
distribution, demonstrated for iid variables in Sec. II, continues to hold for
correlated signals, which are time-translationally invariant. This is then used
to determine the scaling of the singularities of the
two equivalent distributions. Finally, Sec.~VI contains concluding remarks.

\section{MRH$_{\rm I}$ Distribution for Independent, Identically Distributed Variables}

Consider $N$ random variables $h_1,\thinspace
h_2,\thinspace\dots,\thinspace h_{_N}$, selected independently
according to the probability density $p(h)$, called the parent
density. The probability that the variable $h_i$ is less than $x$ is
$\mu(x)=\int_{-\infty}^xp(h)\thinspace dh$, so the probability
$M_{\rm max}(x,N)$ that all $N$ variables take values less than $x$ is
\begin{equation}
M_{\rm max}(x,N)=\mu^N(x)\;.\label{Mmax}
\end{equation}
Since $M_{\rm max}(x,N)$ also represents the
probability that $h_{\rm max}=\max_i{h_i}$ is smaller than $x$, the
probability density of the maximum is
\begin{equation} P_{\rm
max}(x,N)=\frac {\partial}{\partial x}\thinspace M_{\rm
max}(x,N)\;.\label{Pmax1}
\end{equation}

The asymptotic form of the distribution function (\ref{Pmax1}) for large $N$ is
discussed in standard textbooks \cite{gumbel,galambos} on extreme value statistics.
For a wide class of parent distributions, $M_{\rm max}(x,N)$ becomes independent of
$N$ in the large $N$ limit, on making the linear change of variable
$x=a_{_N}z+b_{_N}$ with suitable parameters $a_{_N}$, $b_{_N}$:
\begin{equation}
M_{\rm max}(a_{_N}z+b_{_N},N)\to M_{\rm max}^\ast(z)\;,\quad P_{\rm max}^\ast(z) =
\frac{d}{dz} M_{\rm max}^\ast(z)\;.\label{Mmax(z)}
\end{equation}
Depending on the tail of the parent distribution, the limit function $M_{\rm
max}^\ast(z)$ belongs to one of three classes, associated with the names of
Fisher-Tippett-Gumbel, Fr\'echet, and Weibull. \cite{ft,gn,gumbel,galambos}.

We now turn to the main subject of this paper, the statistics of the quantity
$h_{\rm{max}} - h_1$, i.e., of the maximum height relative to the initial
height. For identically distributed variables it does not matter which $h_i$ is
singled out as a reference, but to be specific we use $h_1$. Let us calculate the
probability ${\cal F}(m, N)$ that all $N$ of the relative variables $h_1-h_1,\thinspace
h_2-h_1,\thinspace\dots,\thinspace h_N-h_1$ are less than $m$. Since the first of
these relative variables is identically zero, ${\cal F}(m,N)$ vanishes for negative $m$ and
for positive $m$ is the same as the probability that
$h_2-h_1,\thinspace\dots,\thinspace h_N-h_1$ are all less than $m$. Thus,
\begin{eqnarray} {\cal F}(m,N)&=&\Theta(m)
\int_{-\infty}^\infty dh_1 p(h_1)\int_{-\infty}^{h_1+m} dh_2
p(h_2)\dots\int_{-\infty}^{h_1+m} dh_{_N} p(h_{_N})\;,\nonumber\\
&=&\Theta(m)\int_{-\infty}^\infty dh_1\thinspace p(h_1)M_{\rm
max}(h_1+m,N-1)\;,\label{M(m,N)}
\end{eqnarray}
where $\Theta(x)$ is the standard Heaviside function, and in going from the first
line to the second we have used Eq. (\ref{Mmax}). Differentiating ${\cal F}(m,N)$ in Eq.
(\ref{M(m,N)}) with respect to $m$ and making use of Eq. (\ref{Pmax1}) we obtain
\begin{equation} P(m,N)={1\over
N}\thinspace\delta(m)+\Theta(m)\int_{-\infty}^\infty dh \, p(h)\,
P_{\rm max} (h+m,N-1)\;, \label{P(m,N)}
\end{equation}
for the probability density of the maximum of all $N$ relative variables.

To analyze the asymptotic form of $P(m,N)$ for large $N$, we neglect
the first term on the right-hand side of Eq. (\ref{P(m,N)}) and
substitute the limiting distribution
\begin{equation} P_{\rm max}
(x,N-1) \approx a_{_N}^{-1} P^\ast_{\rm max}
(a_{_N}^{-1}(x-b_{_N}))\;,\label{limPmax2}
\end{equation}
introduced
in Eq. (\ref{Mmax(z)}), in the second term. The convolution of the two
normalized distribution functions $P^\ast$ and $p$ in
Eq. (\ref{P(m,N)}) depends on their relative scales, and we
distinguish the three cases: (i) $a_{_N}$ vanishes, (ii) $a_{_N}$
converges to a finite $a$ value, and (iii) $a{_N}$ diverges. In the
first case $P_{\rm max} (x,N-1)\approx \delta(x-b_{_N})$, whence
$P(m,N) \approx p(m-b_{_N})$. In case (ii), $P_{\rm max}$ and
$p$ vary on the same scale, and the MRH$_{\rm I}$ distribution
is a convolution of two non-degenerate functions.  As we shall see
later, in this case $P_{\rm max}^\ast(z)$ has the
Fisher-Tippett-Gumbel form, $P_{\rm
max}^\ast(z)=\exp\left(-z-e^{-z}\right)$.  Finally, in case (iii),
$a_{_N} p(a_{_N}z)\approx \delta(z)$, and the integral is readily
evaluated on changing the integration variable to $z$. The results can
be summarized as
\begin{equation}
P(m,N)\approx\left\{\begin{array}{l}p\left(b_{_N}-m\right)\;,\quad
\\ R(m-b_{_N}),\quad\\ a_{_N}^{-1}P^\ast_{\rm
max}\left(a_{_N}^{-1}\left(m-b_{_N}\right)\right)\;,
\quad\end{array}\right.
\begin{array}{l}a_{_N}\to 0\\a_{_N}\to
a\\a_{_N}\to\infty
\end{array} \label{P(m,N)largeN}
\end{equation}
where
\begin{equation} R(x)= \int_{-\infty}^\infty
dz\thinspace p(az-x)\exp\left(-z-e^{-z}\right)\;.  \label{R-fcn}
\end{equation}

We now review the connection between the scale parameters $a_{_N}$, $b_{_N}$,
which play such an important role here, and the parent distribution
\cite{ft,gn,gumbel,galambos}. It is useful to work with the functions $g(x)$ and
$f(z)$, defined by
\begin{equation}
\mu(x)=\exp\left[-e^{-g(x)}\right]\;,\quad M_{\rm
max}^\ast(z)=\exp\left[-e^{-f(z)}\right].\label{g&f-def}
\end{equation}
Here $\mu(x)$ is the integrated parent distribution
introduced above Eq.  (\ref{Mmax}), and $M_{\rm max}^\ast(z)$ is the
limit function defined in Eq.  (\ref{Mmax(z)}). Note that $g$ diverges
as $\mu$ approaches $1$, and its asymptotic form is related to the
tail of the parent distribution. According to Eqs.  (\ref{Mmax(z)})
and (\ref{g&f-def}),
\begin{equation} P_{\rm max}^\ast(z)={d\over
dz}\thinspace\exp\left[-e^{-f(z)}\right]\;.\label{Pmax3}
\end{equation}

In terms of $g(x)$ and $f(z)$, the $N$-independence
(\ref{Mmax(z)}) of the extreme distribution for large $N$ takes the
form
\begin{equation} g\left(a_{_N}z+b_{_{N}}\right)-\ln N\to
f(z)\;.\label{gandf}
\end{equation}
We define the scale factors $a_{_N}$ and $b_{_N}$ by
\begin{equation}
g(b_{_N})= \ln N\;,\quad g^\prime(b_{_N})= a_{_N}^{-1}\;.\label{bNaN}
\end{equation}
Here the conditions $f(0)=0$ and $f^\prime(0)=1$ have been imposed for convenience,
but may be relaxed by making a linear transformation $z=\alpha z'+\beta$, with
parameters $\alpha$ and $\beta$ which are independent of $N$. Such transformations
do not change the asymptotic behavior of $a_{_N}$ and $b_{_N}$.

We now relate the three cases in Eq. (\ref{P(m,N)largeN}) to the tail of the parent
distribution. According to (\ref{bNaN}),
\begin{equation}
a_{_N}=\frac{d\,b_N}{d\ln N}\;,\label{aN}
\end{equation}
which implies that $a_N$ (i) goes to zero, (ii) converges to a constant, or (iii)
diverges if the asymptotic dependence of $b_N$ on $\ln N$ is slower than linear,
linear, or, faster than linear. From Eq. (\ref{bNaN}) we see that in these three
cases $g(x)$ diverges, for $x\to\infty$, (i) faster than linearly,
(ii) linearly, or, (iii) slower than linearly, and since $\mu(x)\approx 1-e^{-g(x)}$ for
large $g$,
$1-\mu(x)$ vanishes (i) faster than exponentially,
(ii) exponentially, or (iii) slower than exponentially, respectively.
Here ``exponentially" is understood in the broad sense as
corresponding to a linear leading divergence of $g(x)$.
Case (ii) therefore includes parent
distributions for which $1-\mu(x)$ decays as $x^\Delta e^{-cx}$ or
$e^{bx^\epsilon}\, e^{-cx},\,\, \epsilon<1$, since, for all these cases,
$g(x)\approx cx$ to leading order as $x\to\infty$.

Finally, recalling that $p(x)=d\mu(x)/dx$, we conclude that the MRH$_{\rm I}$
distribution is given by the second line on the right-hand side of Eq.
(\ref{P(m,N)largeN}) whenever the parent density distribution $p(x)$ decays
exponentially for large $x$, in the broad sense just defined. In this case
$P^\ast_{\rm max}$ is of Fisher-Tippett-Gumbel form as used in
Eq.~\ref{P(m,N)largeN}. If the decay is more
rapid than exponential or less rapid than exponential, then lines one and three in
Eq. (\ref{P(m,N)largeN}) apply.

It is enlightening to supplement this general discussion with explicit results for
some characteristic parent distributions. We begin with the
case of a parent distribution $p(x)$ which for large $x$ decays according to the
generalized exponential form $x^\Delta\exp\left[-(x/\xi)^\delta\right]$, where
$\delta>0$ but $\Delta$ can have either sign. According to Eq. (\ref{g&f-def}),
$g(x)$ has the asymptotic form $g(x)\approx (x/\xi)^\delta$ for
large $x$ to leading
order, independent of $\Delta$, and from Eqs. (\ref{gandf}) and (\ref{bNaN}),
\begin{eqnarray}
&&f(z)=z\;,\quad -\infty<z<\infty\;,\label{fFTP}\\
&& a_{_N}\approx\xi\delta^{-1}(\ln N)^{1/\delta-1}\;,\quad b_{_N}\approx\xi(\ln
N)^{1/\delta}\;.\label{aNbNFTP}
\end{eqnarray}
From Eq. (\ref{aNbNFTP}), we see that $a_{_N}\to 0,\;a$, and $\infty$ for
$\delta>1$, $\delta=0$, and $\delta<1$, respectively. Thus, the corresponding
MRH$_{\rm I}$ distributions are given by the first, second, and third lines,
respectively, on the right side Eq. (\ref{P(m,N)largeN}), in agreement with the
general conclusions of the preceding paragraph. Substituting the scaling function
(\ref{fFTP}) in Eq. (\ref{Pmax3}) yields the Fisher-Tippett-Gumbel form of the
extreme distribution $P_{\rm max}^\ast(z)$, already shown explicitly just above Eq.
(\ref{P(m,N)largeN}) and in Eq. (\ref{R-fcn}).

Next we consider a parent $p(x)$ which vanishes for $x$ greater than a finite
value $x_1$ and varies as $(x_1-x)^{\alpha-1}$, with $\alpha>0$, as $x$ approaches
$x_1$ from below, so that $1-\mu(x)\sim(x_1-x)^\alpha$, and $g(x)\approx
-\alpha\ln(x_1-x)$ for small $x_1-x$. From Eqs. (\ref{gandf}) and (\ref{bNaN}),
\begin{eqnarray} &&f(z)=-\alpha\ln\left(1-\alpha^{-1}z\right)\;, \quad
0<z<\alpha\;,\label{fWeibull}\\ && a_{_N}\approx\alpha^{-1}N^{-1/\alpha}\;,\quad
b_{_N}\approx x_1-N^{-1/\alpha} \;.\label{aNbNWeibull}
\end{eqnarray}
Since $a_N\to 0$ in the limit $N\to\infty$, the MRH$_{\rm I}$ distribution is given
by the top line on the right side Eq. (\ref{P(m,N)largeN}). This is consistent with
the general analysis given above, since the parent distribution has a faster than
exponential decay for $x\to\infty$, having already attained 0 at the finite value
$x_1$. Substituting the scaling function (\ref{fWeibull}) in Eq. (\ref{Pmax3})
yields the Weibull form of the extreme distribution $P_{\rm max}^\ast(z)$.

Finally we consider a parent distribution $p(x)$ which for large $x$ decays as
$x^{-1-\beta}$, with $\beta>0$, so that $1-\mu(x)\sim x^{-\beta}$. In this case
$g(x)\approx \beta\ln x$, and from Eqs. (\ref{gandf}) and (\ref{bNaN}),
\begin{eqnarray} &&f(z)=\beta\ln\left(1+\beta^{-1}z\right)\;,\quad
-\beta<z<\infty\;,\label{fFrechet}\\ &&
a_{_N}\approx\beta^{-1}N^{1/\beta}\;,\quad b_{_N}\approx
N^{1/\beta}\;,\label{aNbNFrechet}
\end{eqnarray}
Since $a_N\to 0$, the MRH$_{\rm I}$ distribution is given by the third line on the
right side Eq. (\ref{P(m,N)largeN}). This is also consistent with the general
analysis given above, since the parent distribution has a slower than exponential
decay. Substituting the scaling function (\ref{fFrechet}) in Eq. (\ref{Pmax3})
yields the Fr\'echet extreme distribution.

The expressions for $f(z)$ in Eqs. (\ref{fFTP}), (\ref{fWeibull}), and
(\ref{fFrechet}), correspond to the special cases $\eta=0$, $\eta<0$ and $\eta> 0$,
respectively, of the generalized extreme value distribution
\cite{{DeHaanResnick:1996}} with
$f(z)=\eta\ln(1+\eta^{-1} z)$. Here we have treated the
three cases separately in order to highlight the different asymptotes
of $a_N$ and $b_N$.

Recently, the density of near extreme states, corresponding to the distribution of
the relative variables $h_{\rm max}-h_i$ and defined by
\begin{equation} \rho (r,N)={1\over N}\sum_{h_i\ne h_{\rm
max}}^{N-1}\delta\left[r-(h_{\rm max}-h_i)\right]\;,\label{sabmaj}
\end{equation}
was studied by Sabhapandit and Majumdar \cite{sm} for
iid variables. Since the MRH$_{\rm I}$ distribution
$P(m,N)=\langle\delta\left([m-(h_{\rm max}-h_1)\right]\rangle$ that we
consider does not depend on the particular variable $h_i$ chosen as a
reference, it may be rewritten as
\begin{equation} P(m,N)={1\over
N}\sum_{i=1}^N\delta\left[m-(h_{\rm
max}-h_i)\right]=N^{-1}\delta(m)+\rho (m,N)\;.\label{MRHI2}
\end{equation}
Thus, the density of near extreme states $\rho (m,N)$ and the MRH$_{\rm I}$
distribution $P(m,N)$ only differ by a delta function. The results of Sabhapandit
and Majumdar for the limiting behavior of $\rho (m,N)$ for large $N$ are essentially
the same as ours for $P(m,N)$, apart from our more general evaluation
of the threshold case of exponentially decaying parent.

In Section VI we point out that the equivalence between the density of near extreme
states and the MRH$_{\rm I}$ distribution is not limited to iid variables, but also
holds for the periodic, correlated signals considered in Sec. III-VI.

\section{Correlated Gaussian Signals}
\label{1overfalpha}
We now turn to Gaussian signals of periodicity $h(t) = h(t+T)$ with configurational
weight \cite{gyorgyietal,antaletal}
\begin{equation}
  \mathcal{P} [h(t)] \propto e^{-S[h(t)]}\;,\label{weight}
\end{equation}
where the effective action, in Fourier space, is
\begin{equation}
  S[c_k;\alpha] = (2\pi)^{\alpha} T^{1-\alpha} \sum_{n=1}^{N/2}
  n^{\alpha}\label{Fourieraction}
  |c_n|^2\;.
\end{equation}
Here the $c_k$ are coefficients in the finite Fourier series
\begin{equation}
h(t) = \sum_{n=-N/2+1}^{N/2} c_n e^{2\pi i nt/T}\;,\quad
c_n^*=c_{-n}\;,\label{Fourierseries}
\end{equation}
where $N$ is a positive, even integer. Since the maximum frequency appearing in the
sum is of order $N/T$, the series does not resolve fine structure on a time scale
less than $\tau = T/N$.

We will be mainly interested in the continuum limit $N\to\infty$, $\tau\to 0$
with $T=N\tau$ fixed. Expressed in terms of $h(t)$ instead of its
Fourier transform, the action in Eqs. (\ref{weight}) and
(\ref{Fourieraction}) takes the form
\begin{equation}
S[h(t)]={1\over 2}\int_0^T dt\thinspace\left\vert{d^{\alpha/2}h\over
dt^{\alpha/2}}\right\vert^2 \;,\label{realspaceaction}
\end{equation}
in this limit, which implies the stochastic equation of motion
\begin{equation}
{d^{\alpha/2}h\over dt^{\alpha/2}}=\xi(t)\;,\quad \langle\xi(t)\xi(t')\rangle=
\delta(t-t')\;,\label{stochasticeq}
\end{equation}
where $\xi(t)$ is Gaussian white noise with zero mean.

The requirement $c_n*=c_{-n}$ in Eq. (\ref{Fourierseries}) guarantees that $h(t)$ is
real. The Fourier coefficients $c_0$ (present for $N$ even only) and $c_{_{N/2}}$
are real, but the other $c_n$ are complex. Configurational averages involve
integration with the statistical weight (\ref{weight}), (\ref{Fourieraction}) over
the phase space
\begin{eqnarray}
\int_{-\infty}^\infty dc_{_{N/2}}\prod_{k=1}^{N/2-1}\int_{-\infty}^\infty
d\thinspace{\rm Re}\left[c_k\right] \int_{-\infty}^\infty d\thinspace{\rm
Im}\left[c_k\right]\;.\label{phasespace}
\end{eqnarray}

From Eqs. (\ref{weight}) and (\ref{Fourieraction}) one sees that the amplitudes of
the Fourier modes are independent, Gaussian distributed variables, but only for
$\alpha=0$ are they identically distributed. This is also apparent from the mean
square amplitude $\langle |c_n|^2 \rangle \propto n^{-\alpha}$, which is consistent
with a $1/f^\alpha$ power spectrum and independent of $n$ only for $\alpha=0$.
Tuning $\alpha$ allows us to treat a broad range of time signals and recover some
important special cases. The values $\alpha = 0,1,2,4$ correspond, respectively, to
white-noise (iid Gaussian variables), $1/f$ noise \cite{mbw}, the random walk
(diffusion), and the random acceleration process \cite{twb93}. For $\alpha=0,2,4$
this correspondence is immediately apparent from the stochastic equation of motion
(\ref{stochasticeq}).

Although the Fourier components $c_n$ are uncorrelated, the corresponding time
signal $h(t)$ is correlated at different times $t$ and $t'$ for $\alpha>0$, and the
correlation increases with increasing $\alpha$. For example, for $\alpha = 0$ the
$h(t)$ at different times are iid random variables, whereas for $\alpha \to \infty$,
$h(t)$ becomes a single-mode sinusoidal curve, as discussed below. For $0 \le \alpha
< 1$, the correlation function $\langle h(t')h(t'+t)\rangle$ is bounded, while for
$\alpha > 1$ it diverges in the limit $T\to\infty$ with $t/T$ finite. For a more
detailed discussion of correlations in $1/f^{\alpha}$ signals, see
\cite{gyorgyietal}.

In the next two Sections we study the distribution function of $m={\rm
max}_t[h(t)-h(0)]$, i.e., of the maximum height with respect to the initial
height (MRH$_{\rm I}$), for the Gaussian model defined by Eqs. (\ref{weight}) and
(\ref{Fourieraction}). First we derive the MRH$_{\rm I}$ distribution exactly in the special
cases $\alpha=0$, 2, 4, and $\infty$ and then, for other, intermediate values of
$\alpha$ determine the distribution with numerical simulations.

In our simulations the distribution $P(m,N)$ of $m$ was computed from about $10^6$
to $10^7$ independent signals $h(t)$ for each of the various values of $N$ and
$\alpha$ that were considered. Each signal was generated by selecting the Fourier
coefficients $c_{_{-N/2+1}}\dots,c_{_{N/2}}$ randomly from the Gaussian distribution
(\ref{weight}), (\ref{Fourieraction}) and then summing the series
(\ref{Fourierseries}) to obtain $h(t)$ at $t=0,\thinspace T/N,\thinspace
2T/N,\thinspace,\dots,(N-1)T/N$. For each signal the maximum $m$ of these $N$
heights relative to the initial height $h(0)$ was determined, and then the values
of $m$ for the $10^6$ to $10^7$ independent signals were binned to obtain the
distribution $P(m,N)$. This procedure was carried out for increasingly large
values of $N$, to obtain the best estimate of the distribution in the
continuous time limit
$N\to\infty$, $\tau\to 0$ with $T=N\tau$ fixed.

To extract scaling functions $\Phi(x)$, free of fitting parameters, from $P(m,N)$ in
the limit $N\to\infty$, we follow the same procedure as Gy\"orgyi et al.
\cite{gyorgyietal}. In cases where $\langle m\rangle_{_N}=\int_0^\infty mP(m,N)dm$
and $\sigma_{_N}=\langle\left(m-\langle m\rangle_{_N}\right)^2\rangle_{_N}^{1/2}$
have the same large $N$ behavior, i.e. where their ratio approaches a constant for
large $N$, we scale by the average, introducing the variable
\begin{equation}
x={m\over\langle m\rangle_{_N}}\label{x}
\end{equation}
and defining the rescaled distribution by
\begin{equation}
\Phi(x)=\lim_{_{N\to\infty}} \langle m\rangle_{_N}P(\langle m\rangle_{_N}x,N)
 \;.\label{Phi}
\end{equation}
Since $m={\rm max}_t[h(t)-h(0)]$ is non-negative, the distribution function
$\Phi(x)$ is only defined for $x\ge 0$. According to Eqs. (\ref{x}) and (\ref{Phi}),
$\Phi(x)$ is normalized so that $\int_0^\infty dx\thinspace\Phi(x)=1$ and has the
mean value $\langle x\rangle=\int_0^\infty dx\thinspace x \Phi(x)=1$.

If, on the other hand, $\langle m\rangle_{_N}$ and $\sigma_{_N}$ have different
large $N$ behavior, we  scale by the standard deviation ($\sigma$ scaling),
introducing the variable
\begin{equation}
y = {m - \langle m \rangle_{_N}\over\sigma_{_N}}\;\label{y}
\end{equation}
and defining the corresponding distribution function by
\begin{equation}
 \tilde\Phi(y)=\lim_{_{N\to\infty}}
 \sigma_{_N} P(\langle m\rangle_{_N}+\sigma_{_N}y,N)\;.\label{Phitilde}
\end{equation}
According to Eqs. (\ref{y}) and (\ref{Phitilde}), $\tilde{\Phi}(y)$ is
normalized so
that $\int_{-\infty}^\infty dy\thinspace\tilde{\Phi}(y)=1$ and has the moments
$\langle y\rangle=\int_{-\infty}^\infty dy\thinspace y\tilde{\Phi}(y)=0$,
$\langle\left(y-\langle y\rangle\right)^2\rangle^{1/2}=1$.

\section{Exact results}
\subsection{Special case $\alpha =0$}\label{alpha=0}
As mentioned below Eq. (\ref{phasespace}), the correlated Gaussian signals $h(t)$
defined in the preceding Section reduce to iid variables in the limit $\alpha\to 0$.
Since the Gaussian parent distribution has a faster than than exponential decay, the
MRH$_{\rm I}$ distribution, given by the upper entry in Eq. (\ref{P(m,N)largeN}), is
also Gaussian. Since $\langle m\rangle_N \sim (\ln N)^{1/2}$ and $\sigma_N \sim (\ln
N)^{-1/2}$ for large $N$, we specify the distribution in terms of the variable $y$
and the function $\tilde{\Phi}(y)$ of $\sigma$ scaling, defined in Eqs. (\ref{y})
and (\ref{Phitilde}):
\begin{equation}
\tilde{\Phi}_{\rm I}(y)=(2\pi)^{-1/2}\exp(-y^2/2)\;,\quad
-\infty<y<\infty\;.\label{scalefuncalpha=0}
\end{equation}

\begin{figure}[htb]
\includegraphics[width=8cm]{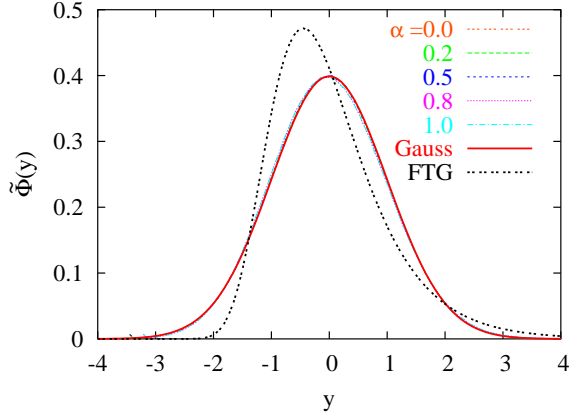}
\caption{Results for the MRH$_{\rm I}$ distribution from simulations, plotted using
$\sigma$-scaling (Eqs.(\ref{y}) and (\ref{Phitilde})), for $\alpha$ in the range
$0\le \alpha \le 1$. The simulations were performed with $N=4096$ terms in the
Fourier series (\ref{Fourierseries}). The exact analytical result
(\ref{scalefuncalpha=0}) for $N\to\infty$ (normalized Gaussian) is also shown, and
the simulation results are within the linewidth of the Gaussian. The curve FTG
corresponds to the Fisher-Tippett-Gumbel distribution, which is the limiting
MRH$_{\rm A}$ distribution in the interval $0\le \alpha <1$.} \label{F:a0-1}
\end{figure}

In contrast to Eq. (\ref{scalefuncalpha=0}), the limiting MRH$_{\rm A}$
distribution, considered in Ref. \cite{gyorgyietal}, has the Fisher-Tippett-Gumbel
form given by Eqs. (\ref{Pmax3}) and (\ref{fFTP}) and shown explicitly just above
Eq. (\ref{P(m,N)largeN}). The two distributions are compared in Fig.\ref{F:a0-1}.

The simulation results shown in Fig.~\ref{F:a0-1} will be discussed in the next
Section. It turns out that the MRH$_{\rm I}$ distribution has the Gaussian form
(\ref{scalefuncalpha=0}) not just for $\alpha=0$, but throughout the interval $0\le
\alpha \le 1$.

\subsection{Special case $\alpha =2$}

As mentioned after Eq.\ (\ref{phasespace}), the correlated Gaussian signals defined
in the preceding Section correspond, for $\alpha=2$, $N\to\infty$, to random walks
$h(t)$ governed by the stochastic equation of motion (\ref{stochasticeq}). The
statistical weight or propagator $Z(h,h_0,t)$ for a random walk from initial
position $h_0$ to $h$ in a time $t$ satisfies the diffusion equation
$\left(\partial/\partial t-D\thinspace\partial^2/\partial h^2\right)Z=0$, with
initial condition $Z(h,h_0,0)=\delta(h-h_0)$. We set $D={1 /2}$ from now on,
corresponding to Gaussian signals normalized as in the preceding Section. The
particular value of $D$ is not important, as it drops out of the MRH$_{\rm I}$
distribution on scaling by the average, as in Eqs. (\ref{x}) and (\ref{Phi}).

We will need the well known solutions of the diffusion equation
\begin{equation}
Z_0(h,h_0,t)={1\over(2\pi t)^{1/2}}\exp\left[-(h-h_0)^2/2t\right]\;\label{propagrw1}
\end{equation}
for random walks in the unbounded space $-\infty<h<\infty$, and
\begin{equation}
Z_1(h,h_0,t)=Z_0(h,h_0,t)-Z_0(h,-h_0,t)\label{propagrw2}
\end{equation}
for random walks in the half space $h>0$, with an absorbing boundary
\cite{absorbingboundary} at $h=0$.

Let us consider the family of random walks $h(t)$ in the unbounded space
$-\infty<h<\infty$ which satisfy the periodic boundary condition $h(0)=h(T)$.
With
no loss of generality in the result for the MRH$_{\rm I}$ distribution,
one may choose
$h(0)=0$. The fraction ${\cal F}_{\rm per}(m,T)$ of the walks with endpoints
$h(0)=h(T)=0$ which never exceed height $m$ in the interval $0<t<T$ may be expressed
as
\begin{eqnarray}
{\cal F}_{\rm per}(m,T)&=&{Z_{h<m}(0,0,T)\over
Z_0(0,0,T)}\nonumber\\&=&{Z_1(m,m,T)\over
Z_0(0,0,T)}=1-e^{-2m^2/T}\;.\label{Fperalpha=2}
\end{eqnarray}
Here $Z_0$ and $Z_1$ are the whole and half-space propagators of Eqs.
(\ref{propagrw1}) and (\ref{propagrw2}), and $Z_{h<m}(h_1,h_0,T)$ is the propagator
for random walks from $h_0$ to $h_1$ in time $T$ which lie entirely in the subspace
$h<m$. In going from the first line of Eq. (\ref{Fperalpha=2}) to the second, we
have used the relation $Z_{h<m}(h_1,h_0,T)=Z_1(m-h_1,m-h_0,T)$, which follows from
invariance of the statistical weight under the coordinate transformation $h\to m-h$.

Differentiating ${\cal F}_{\rm per}(m,T)$ in Eq. (\ref{Fperalpha=2}) with respect to
$m$ yields the MRH$_{\rm I}$ distribution
\begin{equation}
P_{\rm per}(m,T)={4m\over T}\thinspace e^{-2m^2/T}\label{Pperalpha=2}\\
\end{equation}
for random walks with periodic boundary conditions.
The mean value of the MRH$_{\rm I}$ distribution (\ref{Pperalpha=2})
is $\langle m\rangle=(\pi T/8)^{1/2}$. In terms
of the variables $x=m/\langle m\rangle$ and $\Phi(x)=\langle m\rangle
P_{\rm per}(m,T)$ of Eqs. (\ref{x}) and (\ref{Phi}), the distribution
takes the form
\begin{equation}
\Phi_{\rm I}(x)=\frac{\pi}{2}\thinspace x e^{-\pi x^2/4}\;.\label{Phiperalpha=2}
\end{equation}

The mean value $\langle m\rangle$ of the MRH$_{\rm I}$ distribution given in the
preceding paragraph is the same as the mean value $\langle h_m\rangle$ of the
MRH$_{\rm A}$ distribution for $\alpha=2$ obtained by Majumdar and Comtet \cite{mc}.
The equality
\begin{equation}
\langle m\rangle=\langle h_m\rangle\label{equalmeans}
\end{equation}
is a general consequence of the definitions $m={\rm max}_t[h(t)-h(0)]$, $h_m={\rm
max}_t[h(t)-\overline h]$, where $\overline h$ is the time average of $h(t)$, and
time-translational invariance,  on averaging over all paths, in the form $\langle
h(0)\rangle =\langle \overline h\rangle$. Note that the equality (\ref{equalmeans})
holds for all $\alpha$, not just $\alpha=2$, and for non-periodic as well as
periodic boundary conditions, as long as the boundary conditions
are consistent with the time translational invariance.

\begin{figure}[htb]
\includegraphics[width=8cm]{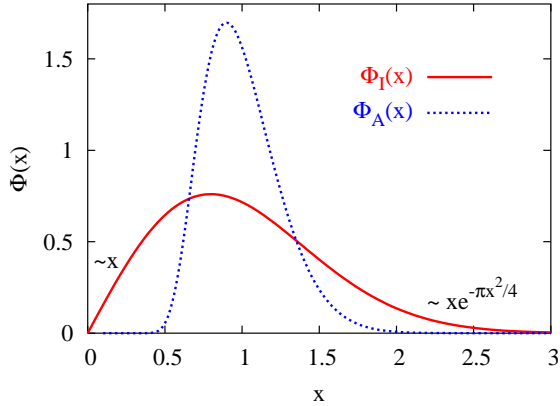}
\caption{The exact MRH$_{\rm I}$ distribution (\ref{Phiperalpha=2}) for $\alpha=2$
and periodic boundary conditions, and the corresponding MRH$_{\rm A}$ or Airy
distribution \cite{mc}. Both distributions are scaled with the average, as in Eqs.
(\ref{x}) and (\ref{Phi}).  } \label{F:a2.0}
\end{figure}

As in the case $\alpha=0$, the MRH$_{\rm I}$ and MRH$_{\rm A}$ distributions for
$\alpha=2$ are not the same. Majumdar and Comtet \cite{mc} have shown that for
random walks with periodic boundary conditions, the MRH$_{\rm A}$ distribution is
the so-called Airy distribution. The two distributions are compared in Fig.
\ref{F:a2.0}. We see that, for both small and large $x$, $\Phi_{\rm I}(x)$ has the
greater weight. These features are found for all $\alpha>1$ and can be understood
heuristically as follows:

In the case of $\Phi_{\rm A}(x)$, small $x$ means small $h_m=h_{\rm max}-
\overline{h}$, or $h_{\rm max}\approx\overline{h}$. There are very few such
configurations, only those for which $h(t)$ is nearly constant. In the case of
$\Phi_{\rm I}(x)$, on the other hand, small $x$ means small $m=h_{\rm max}-h(0)$, or
$h_{\rm max}\approx h(0)$. This condition is far less restrictive, since it is
satisfied by any configuration $h(t)$ which is below $h(0)$ most of the time. Thus,
small $m$ is much more probable than small $h_m$, i.e., $\Phi_{I}(x)\gg \Phi_{\rm
A}(x)$ for small $x$.

Turning to the large $x$ behavior, we note that for a configuration $h(t)$ which
makes a very large positive excursion with respect to the initial height $h(0)$,
the average height $\overline{h}$ also tends to be much larger than $h(0)$.
Thus, for such a configuration $m\gg h_m$. Roughly speaking, this means
that a large value of $h_m$ has the same probability as a much larger value
of $m$. This, together with
a probability distribution that decreases rapidly with increasing $m$ implies
$\Phi_{\rm I}(x)\gg \Phi_{\rm A}(x)$ for large $x$, as seen in Fig. \ref{F:a2.0}.

The distribution of the maximum height not only depends on the
reference height from
which it is measured. It also depends on the boundary conditions imposed on $h(t)$
at $t=0$ and $t=T$. Before leaving the special case $\alpha=2$, we derive the
MRH$_{\rm I}$ distribution for random walks for two non-periodic boundary conditions
of general interest.

Consider the family of random walks $h(t)$ on the infinite interval
$-\infty<h<\infty$ with fixed endpoints $h(0)=0$ and $h(T)=h_1$. For this
fixed boundary
condition Eqs. (\ref{Fperalpha=2}) and (\ref{Pperalpha=2}) are replaced by
\begin{eqnarray}
{\cal F}_{\rm fix}(m,h_1;T)&=&{Z_{h<m}(h_1,0,T)\over
Z_0(h_1,0,T)}\nonumber\\&=&{Z_1(m-h_1,m,T)\over Z_0(h_1,0,T)}=
1-\exp\left[{h_1^2-(2m-h_1)^2\over 2T}\right]\;.\label{Fh1alpha=2}
\end{eqnarray}
and
\begin{equation}
P_{\rm fix}(m,h_1;T)={2(2m-h_1)\over T}\thinspace\exp\left[{h_1^2-(2m-h_1)^2\over
2T}\right]\;,\label{Ph1alpha=2}
\end{equation}
which reduces to Eq. (\ref{Pperalpha=2}) for $h_1=0$.

Finally, we consider the family of random walks $h(t)$, $0<t<T$, with initial
condition $h=0$ at $t=0$ but with no restrictions on $h(T)$. For this boundary
condition Eqs. (\ref{Fperalpha=2}) and (\ref{Pperalpha=2}) are replaced by
\begin{eqnarray}
{\cal F}_{\rm free}(m,T)&=&{\int_{-\infty}^m dh_1\thinspace Z_{h<m}(h_1,0,T)\over
\int_{-\infty}^\infty dh_1\thinspace Z_0(h_1,0,T)}=\int_{-\infty}^m dh_1\thinspace
Z_1(m-h_1,m,T)\label{first}\\&=&\int_0^\infty dh\thinspace Z_1(h,m,T)= {\rm
erf}\left({m\over \sqrt{2T}}\right)\;,\label{Fallh1alpha=2}
\end{eqnarray}
where ${\rm erf}$ denotes the error function \cite{as}. The quantity ${\cal F}_{\rm
free}(m,T)$ is the probability that a random walk which begins at the origin has not
yet reached point $m$ after a time $T$. The integral on the right-hand side of Eq.
(\ref{Fallh1alpha=2}) is the ``persistence" probability that a random walk which
begins at point $m$ has not yet reached the origin after a time $T$. Equation
(\ref{Fallh1alpha=2}) states the obvious fact that these two probabilities are equal
and reproduces the well known $T^{-1/2}$ decay of the persistence probability for
long times.

Differentiating ${\cal F}_{\rm free}(m,T)$ in (\ref{Fallh1alpha=2}) with respect to
$m$ yields the MRH$_{\rm I}$ distribution
\begin{equation}
P_{\rm free}(m,T)=\left({2\over\pi T}\right)^{1/2}\thinspace
e^{-m^2/2T}\;.\label{Pallh1alpha=2}
\end{equation}
After scaling by the average, as in Eqs. (\ref{x}) and (\ref{Phi}),
with $\langle m\rangle=(2T/\pi)^{1/2}$, the distribution
function ({\ref{Pallh1alpha=2}) takes the form
\begin{equation}
\Phi_{\rm I}(x)={2\over\pi}\thinspace e^{-x^2/\pi}\;,\label{Phiallh1alpha=2}
\end{equation}
which is clearly different from the result (\ref{Phiperalpha=2}) for periodic
boundary conditions.

\subsection{Special case $\alpha =4$}
\label{S:alpha4}

As pointed out below Eq. (\ref{phasespace}), the correlated Gaussian signal $h(t)$
of Sec.~III may be interpreted, for $\alpha=4$, $N\to\infty$, as the position of a
particle which is randomly accelerated according to the stochastic equation of
motion (\ref{stochasticeq}). Following the same approach as in the preceding
Subsection, we work with the statistical weight or propagator $Z(h,v;h_0,v_0;t)$ for
a randomly-accelerated particle with position and velocity $h_0,v_0$ at $t=0$ and
values $h,v$ at a later time $t$. This quantity satisfies the Fokker-Planck equation
$\left(\partial/\partial t+v\partial/\partial h-{\cal
D}\thinspace\partial^2/\partial v^2\right)Z=0$, which is basically a diffusion
equation for the velocity, with initial condition
$Z(h,v;h_0,v_0;0)=\delta(h-h_0)\delta(v-v_0)$. We set ${\cal D}=1$ from now on, as
in Ref. \cite{twb93}. The particular value of ${\cal D}$ is not important, as it
drops out of the MRH$_{\rm I}$ distribution on scaling by the average, as in Eqs.
(\ref{x}) and (\ref{Phi}).

In calculating the MRH$_{\rm I}$ distribution, we will need the solutions $Z_0$ and
$Z_1$ to the Fokker-Planck equation in the unbounded space $-\infty<h<\infty$ and in
the half space $h>0$, with an absorbing boundary condition \cite{absorbingboundary},
respectively. Expressions for $Z_0(h,v;h_0,v_0;t)$ and for the Laplace transform
$\tilde{Z}_1(h,v;h_0,v_0;s)=\int_0^\infty dt\thinspace e^{-st}Z_1(h,v;h_0,v_0;t)$
are given in Ref. \cite{twb93} and in Appendix A of this paper.

Let us consider the family of trajectories $h(t)$ in the unbounded space
$-\infty<h<\infty$ with periodicity $h(t)=h(t+T)$. As in the case $\alpha=2$, we may
choose $h(0)=0$, with no loss of generality in the result for the MRH$_{\rm I}$ distribution.
The velocities at $t=0$ and $T$ are the same, as follows from the periodicity, but
otherwise unrestricted. All values of the initial velocity are assumed to be equally
probable. For this boundary condition, Eq. (\ref{Fperalpha=2}) in our treatment of
the random walk is replaced by
\begin{eqnarray}
{\cal F}_{\rm per}(m,T)&=&{\int_{-\infty}^\infty dv\thinspace
Z_{h<m}(0,v;0,v;T)\over \int_{-\infty}^\infty dv\thinspace
Z_0(0,v;0,v;T)}\nonumber\\&=&{\int_{-\infty}^\infty dv\thinspace Z_1(m,v;m,v;T)\over
\int_{-\infty}^\infty dv\thinspace Z_0(0,v;0,v;T)}.\label{Fperalpha=4}
\end{eqnarray}
Here $Z_{h<m}(h_1,v_1;h_0,v_0;T)$ is the statistical weight for a randomly
accelerated particle that propagates from point $h_0,v_0$ in phase space to
$h_1,v_1$ in a time $T$ without leaving the half space $h<m$. In going from the
first line of Eq. (\ref{Fperalpha=4}) to the second, we have used the relation
$Z_{h<m}(h_1,v_1;h_0,v_0;T)=Z_1(m-h_1,-v_1;m-h_0,-v_0;T)$, which follows from
invariance of the statistical weight under the coordinate transformation $h\to m-h$. The probability distribution of $m$ is obtained by
differentiating ${\cal F}_{\rm per}(m,T)$ with respect to $m$.

Using the results for $Z_0$ and $Z_1$ in Ref. \cite{twb93}, we have derived the
MRH$_{\rm I}$ distribution for the periodic boundary condition of the preceding
paragraph. The calculation is outlined in Appendix A. For the mean value, one
obtains $\langle m\rangle={1\over 24}\left(\pi T^3\right)^{1/2}$. In terms of the
scaling variable $x$ and scaling function $\Phi(x)$ of Eqs. (\ref{x}) and
(\ref{Phi}), the distribution is given by
\begin{equation}
\Phi_{\rm I}(x)= 2^{-2/3}3^{1/6}\pi^{-1/6}\thinspace x^{-1/3}\thinspace
\exp\left(-\frac{\pi}{12}\thinspace x^2\right)U\left(-\frac{1}{6},\frac{2}{3},
\frac{\pi}{12}\thinspace x^2\right)\;,\label{Phiperalpha=4}
\end{equation}
where $U(a,b,z)$ is Kummer's function \cite{as}. The function $\Phi_{\rm I}(x)$
shown in Fig.~\ref{F:a4.0} has the asymptotic forms \cite{as}
\begin{equation}
\Phi_{\rm I}(x)\approx\left\{\begin{array}{l}2^{-4/3}3^{1/6}\pi^{-2/3}
\Gamma({2\over
3})\thinspace x^{-1/3}\;,\label{Phiper4asymp}\\
{1\over 2}\exp\left(-\thinspace{\pi\over
12}x^2\right)\;,\end{array}\right.\quad\begin{array}{l}x\to 0\;,\\
x\to\infty\;,\end{array}
\end{equation}
and the moments
\begin{equation}
\langle x^\nu\rangle={\Gamma({3\over 2}\nu+1)\over\Gamma(\nu+1)}\thinspace
\Gamma(\textstyle{5\over 2})^{-\nu}\;,\label{momentsperalpha=4}
\end{equation}
for arbitrary $\nu >-2/3$.

\begin{figure}[htb]
\includegraphics[width=8cm]{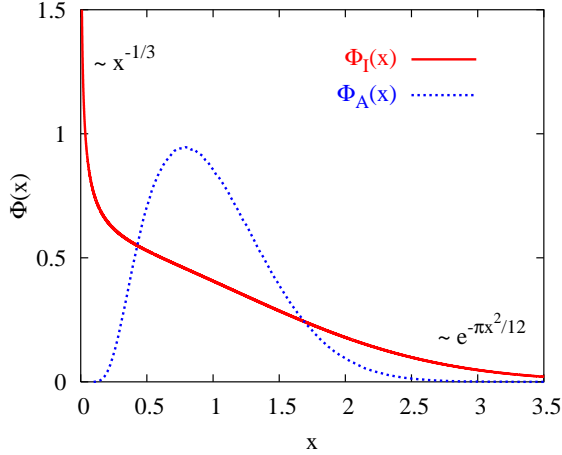}
\caption{The exact MRH$_{\rm I}$ distribution (\ref{Phiperalpha=4}) for $\alpha=4$
and periodic boundary conditions, and the corresponding MRH$_{\rm A}$ distribution,
determined from simulations \cite{gyorgyietal}. Both distributions are scaled with
the average, as in Eqs. (\ref{x}) and (\ref{Phi}). The asymptotic forms of the
MRH$_{\rm I}$ distribution for small and large $x$ are given in Eq.
(\ref{Phiper4asymp}). The MRH$_{\rm A}$ distribution is exponentially small for
small $x$ and for large $x$ is essentially Gaussian with a smaller variance
\cite{gyorgyietal}.} \label{F:a4.0}
\end{figure}

The difference between $\Phi_{\rm I}(x)$ and $\Phi_{\rm A}(x)$ in Fig.~\ref{F:a4.0}
is even more dramatic than for $\alpha =2$. For $\alpha=4$, $\Phi_{\rm I}(x)$
diverges at $x=0$ and decreases monotonically with increasing $x$. The increasing
weight, with increasing $\alpha$, of the MRH$_{\rm I}$ distribution $\Phi_{\rm
I}(x)$ will be discussed below, in connection with simulation results for a broad
range of $\alpha$.

Next, we consider the MRH$_{\rm I}$ distribution for a randomly accelerated particle
with position $h=0$ and velocity $v_0$, but with no restrictions on the position and
velocity at $t=T$. For this boundary condition Eqs. (\ref{first}) and
(\ref{Fallh1alpha=2}) are replaced by
\begin{eqnarray}
{\cal F}_{\rm free}(m,v_0;T)&=&{\int_{-\infty}^m dh_1\int_{-\infty}^\infty
dv_1\thinspace Z_{h<m}(h_1,v_1;0,v_0;T)\over \int_{-\infty}^\infty
dh_1\int_{-\infty}^\infty dv_1\thinspace Z_0(h_1,v_1;0,v_0;T)}\\&=&\int_{-\infty}^m
dh_1\int_{-\infty}^\infty dv_1\thinspace Z_1(m-h_1,-v_1;m,-v_0;T)\\&=&\int_0^\infty
dh\int_{-\infty}^\infty dv\thinspace
Z_1(h,v;m,-v_0;T)=q(m,-v_0;T)\;.\label{Fallh1alpha=4}
\end{eqnarray}
Here ${\cal F}_{\rm free}(m,v_0;T)$ is the probability that a randomly accelerated
particle which begins at the origin with velocity $v_0$ has not yet reached point
$m$ in a time $T$. As in the case $\alpha=2$, ${\cal F}_{\rm free}$ represents a
``persistence" probability. The quantity $q(m,-v_0;T)$ on the right-hand side of Eq.
(\ref{Fallh1alpha=4}) is the probability that a randomly accelerated particle with
initial position $m$ and initial velocity $-v_0$ has not yet reached the origin
after a time $T$. Equation (\ref{Fallh1alpha=4}) states the obvious fact that these
two probabilities are equal and reproduces (see Eq. (\ref{2ndbcalpha=4}) below) the
well-known $T^{-1/4}$ decay \cite{twb93} of the persistence probability for long
times.

Combining $P_{\rm free}(m,v_0;T)=\partial_{\rm free}{\cal F}(m,v_0;T)/\partial m$,
Eq. (\ref{Fallh1alpha=4}), and the expression for the Laplace transform of
$q(m,-v_0;T)$ in Eq. (19) of Ref. \cite{twb93}, we obtain
\begin{eqnarray}
\tilde{P}_{\rm free}(m,v_0;s)&=&\int_0^\infty dT \thinspace e^{-sT}P(m,v_0;T)\\
&=&\int_0^\infty dF\thinspace F^{-2/3}e^{-Fm}{\rm
Ai}\left(-F^{1/3}v_0+F^{-2/3}s\right)\nonumber\\
\qquad &\times&\left[1+{1\over 4\pi^{1/2}}\Gamma\left(-\textstyle{1\over 2},{2\over
3}F^{-1}s^{3/2}\right)\right]\label{Pfreealpha=4}
\end{eqnarray}
for the Laplace transform of the MRH$_{\rm I}$ distribution. Here ${\rm Ai}(z)$ and
$\Gamma(a,z)$ are the standard Airy and incomplete Gamma function. In principle,
$P_{\rm free}(m,v_0,T)$ can be determined from Eq. (\ref{Pfreealpha=4}) by
integrating over $F$ and inverting the Laplace transform numerically. For $v_0=0$
all of the moments $\langle m^\nu\rangle$ of the distribution can be calculated
analytically \cite{twbunpub} from Eq. (\ref{Pfreealpha=4}). Equation (21) of Ref.
\cite{twb93} leads to the exact asymptotic form
\begin{eqnarray} P_{\rm free}(m,v_0;T)&=&
2^{-1}3^{4/3}\pi^{-3/2}\Gamma(\textstyle{1\over
4})\nonumber\\
&\times&{\partial\over\partial m}\left[\left(m^2\over
T^3\right)^{1/12}U\left(-{1\over 6},{2\over 3},\thinspace -{v_0^3\over
9m}\right)\right]\;,\label{2ndbcalpha=4}
\end{eqnarray}
for $T\gg m^{2/3}$ and $T\gg\vert v_0\vert^{1/2}$. Here $U(a,b,z)$ is Kummer's
function \cite{as}.

Finally, we note that the MRH$_{\rm I}$ distribution for $\alpha=4$ and all its
moments can be calculated analytically \cite{twbunpub} for two additional boundary
conditions: In the first case $h(T)=h(0)$ and $v(T)=v(0)=0$. In the second case
$h(T)=h(0)$, while $v(T)$ and $v(0)$ are independent and unrestricted, and
integrated over all values between $-\infty$ and $\infty$.

\subsection{Special case $\alpha=\infty$}
\label{alphainf}

\label{sec:mrh-pi} According to Eqs. (\ref{weight})-(\ref{Fourierseries}),
all but the modes with Fourier coefficients $c_{-1}$ and $c_1$ are
suppressed in the limit $\alpha\to\infty$, so that
\begin{eqnarray}
h(t)-h(0)&=&c_1\left(e^{2\pi i t/T}-1\right)+c_1^*\left(e^{-2\pi i
t/T}-1\right)\\\label{h(t)alpha=01} &=&2\vert c_1\vert\left[\cos\left(2\pi
t/T+\varphi\right)-\cos\varphi\right]\;,\label{h(t)alpha=02}
\end{eqnarray}
where $c_1=\vert c_1\vert e^{i\varphi}$. Thus, $m={\rm max}_t[h(t)-h(0)]=2\vert
c_1\vert(1-\cos\varphi)=4\vert c_1\vert\sin^2(\varphi/2)$, and, in accordance with
Eqs. (\ref{weight}) and (\ref{Fourieraction}), the distribution of $m$ is given by
\begin{eqnarray}
P(m,N)&=&{\lambda\over\pi}\int_0^{2\pi}d\varphi\int_0^\infty d\vert
c_1\vert\thinspace\vert c_1\vert e^{-\lambda\vert c_1\vert^2} \delta\left(m-4\vert
c_1\vert\sin^2{\varphi\over 2}\right)\;,\\
&=&{\lambda m\over 8\pi}\int_0^\pi
d\vartheta\thinspace(\sin\vartheta)^{-4}\exp\left(-{\lambda m^2\over
16\sin^4\vartheta}\right)\;.\label{thetaintegral2}
\end{eqnarray}
Here we have absorbed several constants in $\lambda$, rewritten the integral
$\int_{-\infty}^\infty d\thinspace{\rm Re}\left[c_1\right] \int_{-\infty}^\infty
d\thinspace{\rm Im}\left[c_1\right]$ over phase space (see Eq. (\ref{phasespace}))
in terms of polar coordinates $(\vert c_1\vert$, $\varphi)$, performed the
integration over $\vert c_1\vert$, and made the substitution $\vartheta={1\over
2}\varphi$. By carrying out the $m$ integration before the $\vartheta$ integration,
it is easy to check that $P(m,N)$ in Eq. (\ref{thetaintegral2}) satisfies the
normalization condition $\int_0^\infty dm\thinspace P(m,N)=1$ and has the average
value $\langle m\rangle=(\pi/\lambda)^{1/2}$.

The integral over $\vartheta$ in Eq. (\ref{thetaintegral2}) can be evaluated with
the help of Ref. \cite{gr} or {\it Mathematica}. In terms of the variable
$x=m/\langle m\rangle=(\lambda/\pi)^{1/2}m$ and scaling function $\Phi(x)$ of Eqs.
(\ref{x}) and (\ref{Phi}), the MRH$_{\rm I}$ distribution for periodic boundary
conditions and $\alpha=\infty$ takes the form
\begin{equation}
\Phi_{\rm I}(x)=2^{-3/2}\left({\pi x^2\over 16}\right)^{-1/4}\exp\left(-\thinspace{\pi
x^2\over 16}\right)U\left(-{1\over 4},{1\over 2},{\pi\over 16}\thinspace
x^2\right)\;,\label{PhiSq=infty}
\end{equation}
where again $U(a,b,z)$ is Kummer's function \cite{as}.
The function $\Phi_{\rm I}(x)$ has the asymptotic forms \cite{as}
\begin{equation}
\Phi_{\rm I}(x)\approx\left\{\begin{array}{l}2^{-1}\pi^{-3/4}\Gamma({3\over 4})\thinspace
x^{-1/2}\;,\\
2^{-3/2}\exp\left(-\thinspace{\pi\over
16}x^2\right)\;,\end{array}\right.\quad\begin{array}{l}x\to 0\;,\\
x\to\infty\;,\end{array}\label{Phiinftyasymp}
\end{equation}
and the moments
\begin{equation}
\langle x^\nu\rangle={\Gamma({1\over 2}+\nu)\over\Gamma({1\over 2}+{\nu\over
2})}\thinspace \Gamma(\textstyle{3\over 2})^{-\nu}\;\label{momentsalphainfty}
\end{equation}
for arbitrary $\nu >-{1\over 2}$.

From Eq. (\ref{Phiinftyasymp}) we see that MRH$_{\rm I}$ distribution function
(\ref{PhiSq=infty}) diverges as $x^{-1/2}$ for $x\to 0$, even more strongly than in
the case $\alpha=4$ (see Eq. (\ref{Phiper4asymp})), where there is an $x^{-1/3}$
divergence. The distribution function (\ref{PhiSq=infty}) is plotted in Fig.
\ref{F:a1.2-infty}. The MRH$_{\rm A}$ distribution for $\alpha=\infty$, calculated
by Gy\"orgyi et al. \cite{gyorgyietal}, has precisely the same form
(\ref{Phiperalpha=2}) as the MRH$_{\rm I}$ distribution for $\alpha=2$. Thus, the
curves $\alpha =2$ and $\infty$ in Fig.\ \ref{F:a1.2-infty} correspond to the
MRH$_{\rm A}$ and MRH$_{\rm I}$ distributions, respectively, for $\alpha =\infty$.

\section{Results of simulations}
\label{simu}

In addition to the analytical results for the special values of $\alpha = 0,2,4$ and
$\infty$, we have determined the MRH$_{\rm I}$ distribution for intermediate values
of $\alpha$ by numerical simulations, following the procedure described in
Sec.~\ref{1overfalpha}. This allows us to address some interesting questions
concerning the $\alpha$ dependence. In the simulations the number $N$ of terms in
the Fourier series (\ref{Fourierseries}) was chosen large enough so that finite-size
effects were negligible, and the exact results of the preceding Section,
corresponding to $N=\infty$, could be reproduced with ``width-of-the-line"
accuracy everywhere except in the immediate neighborhood of $x=0$,
where $\Phi_{\rm I}(x)$ diverges
for $\alpha$ greater than a critical value close to 3.

First we turn to the weak-correlation regime $0<\alpha<1$, where the correlation
function $\langle h(0)h(t)\rangle$ remains finite in the limit $T\to \infty$ with
$t/T$ finite, decaying with power law $t^{\alpha -1}$ \cite{gyorgyietal}. Berman
\cite{Berman-1964} has proved that in this regime the correlations are too weak to
affect the MRH$_{\rm A}$ distribution, which has the same FTG form
as for $\alpha=0$. According to the simulation results shown in Fig.~\ref{F:a0-1},
the MRH$_{\rm I}$ distribution has the same Gaussian form (\ref{scalefuncalpha=0})
throughout the interval $0\le\alpha < 1$. Thus, the correlations appear to be
equally irrelevant for the MRH$_{\rm I}$ distribution.

At $\alpha =1$ we enter a regime of stronger correlations. The correlation function
$\langle h(0)h(t)\rangle$ diverges as $\ln T$ at $\alpha=1$ and as $T^{\alpha -1}$
for $\alpha>1$. Although Berman's result on the irrelevance of correlations no
longer holds at $\alpha=1$, the numerically determined MRH$_{\rm I}$ distribution is
still in good agreement with the Gaussian form (\ref{scalefuncalpha=0}).

Once we are well within the strongly correlated regime ($\alpha$ larger than $\sim
1.2$), the convergence with increasing $N$ is much faster than for $\alpha<1$, just
as in the case of the MRH$_{\rm A}$ distribution \cite{gyorgyietal}. Our exact and
numerical results MRH$_{\rm I}$ for $\alpha>1$ are summarized in
Fig.~\ref{F:a1.2-infty}.

\begin{figure}[htb]
\includegraphics[width=8cm]{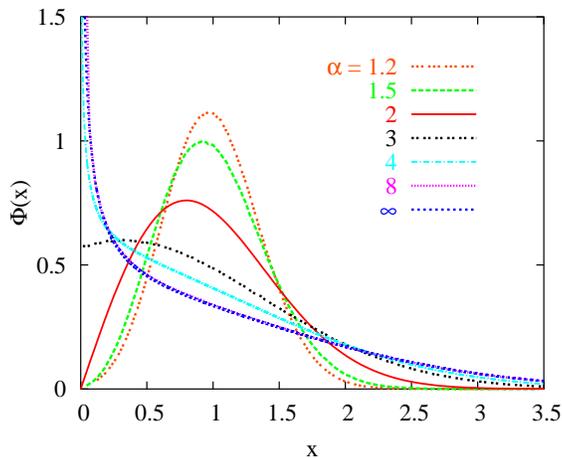}
\caption{MRH$_{\rm I}$ distribution for periodic boundary
conditions and $\alpha$ in
the range $1.2\le \alpha <\infty$, scaled with the average,
as in Eqs. (\ref{x}) and
(\ref{Phi}). For $\alpha=2,\, 4$ and $\infty$ the exact analytic results are shown,
while the results for $\alpha=1.2,\, 1.5,\, 3$ and $8$ were obtained from
simulations with $N=4096$ terms in the Fourier series (\ref{Fourierseries}). Note
that the curves for $\alpha =8$ and $\alpha =\infty$ are practically
indistinguishable.} \label{F:a1.2-infty}
\end{figure}

A prominent feature in Fig. \ref{F:a1.2-infty}, which we have already mentioned, is
the increase in weight, with increasing $\alpha$ of the distribution function
$\Phi_{\rm I}(x)$ for small $x$, with a divergence at $x=0$ for $\alpha$ greater
than a critical value $\alpha_c$. For small $x$,
\begin{equation}
\Phi_{\rm I}(x)\sim x^{\gamma(\alpha)}\;,\quad x\to 0\;,\label{Phigamma}
\end{equation}
with an exponent $\gamma(\alpha)$ that decreases monotonically as $\alpha$
increases, and changes from positive to negative at $\alpha_c$. The Gaussian form of
$P(m,N)$ for $0\le \alpha<1$, with $\langle m\rangle_N\sim\sigma_N^{-1}\sim (\ln
N)^{1/2}$, leads to $\Phi_{\rm I}(x)=\delta(x-1)$ on scaling by the average. This is
consistent with $\gamma(\alpha)=\infty$ for $\alpha<1$ in Eq. (\ref{Phigamma}). The
exact results in Eqs. (\ref{Phiperalpha=2}), (\ref{Phiperalpha=4}), and
(\ref{Phiinftyasymp}) imply $\gamma(2)=1$, $\gamma(4)=-{1\over 3}$, and
$\gamma(\infty)=-{1\over 2}$. The curve for $\alpha=3$ in Fig. \ref{F:a1.2-infty}
looks compatible with $\alpha_c=3$, but determining $\alpha_c$ numerically with good
precision requires a more extensive study, with careful attention to finite size
effects near $x=0$ and $\alpha =3$.

We have also studied the small $x$ singularity for $\alpha = 5$ and $6$ numerically.
In both cases the exponents $\gamma$ are in good agreement with the value $-{1\over
2}$, although the simulations can, of course, not exclude small deviations. These
findings and the exact analytical results $\gamma(4)=-{1\over 3}$ and
$\gamma(\infty)=-{1\over 2}$ appear to indicate that $\gamma(\alpha)=-{1\over 2}$
for all $\alpha>\alpha_u$, where $\alpha_u$ is an upper critical value between 4 and
5. In the next Section we present some simple physical arguments which predict
$\alpha_c=3$ and $\alpha_u =5$. We also obtain a formula for $\gamma(\alpha)$,
$\alpha\le 5$ which reproduces all the known exact values and agrees with the
simulations.

\section{Connection between the MRH$_I$ distribution and the density of states
near extremes} \label{connection}

As we saw at the end of Sec. II, for iid variables the density of states near
extremes, studied by Sabhapandit and Majumdar \cite{sm}, and the MRH$_{\rm I}$
distribution are identical. Here we show that this equivalence continues to hold for
correlated variables, as long as the probability distribution of the signals is
time-translationally invariant. This is the case for correlated Gaussian signals
with $1/f^\alpha$ noise and periodic boundary conditions, and we use the equivalence
with the density of states to study the small $x$ singularity of $\Phi_{\rm I}(x)$.

In a time-translationally invariant system, all of the points traced out by the
signal correspond to possible initial states. For a given realization of the signal,
the distribution of heights measured from the maximum height is the same as the the
distribution of the maximum height with respect to all these possible initial
heights. Thus, averaging over all realizations yields identical distributions for
the quantities $h_{\rm max}-h$ and $h_{\rm max}-h(0)$. We have checked the
equivalence of the two distributions for periodic, correlated Gaussian signals both
numerically and, for $\alpha=$ 2 and 4, by expressing both distributions in terms of
propagators.

We now present a simple picture that is very helpful in understanding the threshold
behavior of the density of states relative to the maximum. The idea is most readily
understood in the $\alpha \to \infty$ limit. In this case, considered in Sec.~\ref{alphainf},
each path is a smooth curve with a parabolic maximum, so that $\delta
h=h(t_m)-h(t)\sim (\delta t)^2$, where $\delta t=t-t_m$. Thus, the density of states
relative to the maximum has the singularity $\rho(\delta h)\sim \delta t/\delta
h\sim(\delta h)^{-1/2}$ for $\delta h\to 0$. Since the density of states is the same
as the MRH$_{\rm I}$ distribution, the exponent $\gamma$ in Eq. (\ref{Phigamma}) has
the value $-{1\over 2}$, in agreement with the exact result in Eq.
(\ref{Phiinftyasymp}).

As long as the path $h(t)$ is smooth near its maximum, in the sense that typical
maxima are parabolic, the above argument applies. Since the average increments of
$1/f^\alpha$ signals scale as
\begin{equation}
|\delta h| \sim \sqrt{\langle \delta h^2 \rangle } \sim |\delta t|^{(\alpha - 1)/2}
 \label{dh-dt}
\end{equation}
for $\alpha>1$, the signals are expected to be twice differentiable for $\alpha\ge
5$. Thus, we predict $\gamma=-{1\over 2}$ for $\alpha\ge \alpha_u=5$, in agreement
with the simulations for $\alpha=$5 and 6.

The lower critical value $\alpha_c$ is the smallest value of $\alpha$ for which the
$1/f^\alpha$ signals are once differentiable, which, according to the scaling
(\ref{dh-dt}) is expected to be $\alpha_c=3$. In this case, the paths near the
maximum are basically rooftop-like, so that $\delta h \sim |\delta t|$ and
$\rho\sim$ constant. This implies $\gamma (3)=0$, which is compatible with the
simulation results for $\alpha=3$  in Fig. \ref{F:a1.2-infty}.

To obtain a formula for $\gamma (\alpha)$ in the interval $1\le \alpha \le 5$, we
assume that the scaling behavior in Eq.~(\ref{dh-dt}) applies near the maximum of
the paths. This implies $\rho(\delta h)\sim \delta t/\delta h\sim (\delta
h)^{(3-\alpha)/(\alpha -1)}$ and so $\gamma (\alpha)=(3-\alpha)/(\alpha -1)$.

In summary, the simple scaling picture predicts the small argument behavior
\begin{equation}
\rho(\delta h)\sim \delta h^\gamma\;,\quad \Phi_{\rm I}(x)\sim
x^\gamma\;,\label{rhoPhi}
\end{equation}
where
\begin{equation}
\gamma(\alpha)=\begin{cases} \displaystyle\frac{3-\alpha}{\alpha -1}\;,
 &\alpha < 5\;,\\\displaystyle
-\frac{1}{2}\;, &\alpha \ge 5 \;, \end{cases}\label{gammaalpha}
\end{equation}
for both the density of near extreme states and the MRH$_{\rm I}$ distribution. The
above derivation may seem oversimplified, but the result is in remarkable agreement
with those presented in Secs. IV and V. Expression (\ref{gammaalpha})
reproduces the exact values of $\gamma(\alpha)$ for $\alpha=$2, 4, and $\infty$
collected below Eq. (\ref{Phigamma}). Furthermore, the divergence
$\gamma\to\infty$ for $\alpha\to 1$ is in accordance with the
fact that, in this limit, the MRH$_{\rm I}$ distribution approaches a
delta function centered at $x=1$.
It also agrees with all our simulations. We
note, however, that a convincing numerical confirmation of $\gamma (\alpha )$ was
not achieved in the regime $1<\alpha\le 1.6$ due to finite-size corrections. We
suspect the relation (\ref{gammaalpha}) is exact, but more effort is needed to put
it on a firmer foundation.

The equivalence of the MRH$_{\rm I}$ distribution and the density of near-extreme
states and the scaling picture presented above furnish a simple physical
explanation for the striking increase of both distributions, for small relative
heights, with increasing $\alpha$. As $\alpha$ increases, high frequency Fourier
components of the Gaussian signal are suppressed. Typical signals become smoother
and approach the maximum height less steeply, remaining close to the maximum for a
longer time. Beyond $\alpha=3$ the approach to the maximum is basically tangential,
becoming parabolic for $\alpha>5$. Thus, the density of near maximum heights $\delta
t/\delta h$ increases as $\alpha$ increases and diverges at the maximum for
$\alpha>3$. A large density of near-maximum heights is equivalent to a high
probability that the maximum height is close to the initial height. The scaling
picture provides both a simple qualitative explanation of the singular
small-argument behavior Eq. (\ref{rhoPhi}) of the two equivalent distributions
$\rho(\delta h)$ and $\Phi_{\rm I}(x)$ and a quantitatively accurate prediction for
the exponent $\gamma(\alpha)$.

\section{Concluding remarks}

In this paper we have shown the importance of the reference point from which the
maximum is measured in extreme statistics. We found that the distributions of the
maximum height relative to the average height and relative to the initial height are generally not the same.
One reason for this is that both the average height and the
initial height are fluctuating quantities, but the distributions of their
fluctuations are different. Furthermore, the two different reference heights
impose different constraints on the paths consistent with a particular
value of the maximum
height. For example, the probability for small maximum height relative to the
average height is small, since only paths which remain close to the
average for the
duration of the signal contribute. In the case of a small height relative to the
initial height, a much larger family of paths, which remain below the initial height most of the time, is allowed. This difference is highlighted
by the analytical result that the MRH$_{\rm A}$ distribution has an
essential singularity at small heights with
exponentially suppressed values \cite{gyorgyietal}, whereas the
MRH$_{\rm I}$ distribution has power law behavior near the origin.

A notable feature of the aforementioned singularity of the MRH$_{\rm I}$
distribution is that the exponent in the power law
monotonically decreases with $\alpha$, thus increasing the
weight near zero heights. This is in agreement with
known persistence
properties of $1/f^\alpha$ processes \cite{Bray-Maj-2001}, where the
weight of configurations persisting below the starting height shows a
similar
trend in $\alpha$. To make this intuitively appealing connection
more rigorous, further considerations would be required.

The small height singularity of the MRH$_{\rm I}$ distribution
was described quantitatively with the help of a remarkable
connection to the density of near extreme states.
Namely, we found that they are identical for
iid variables or, more generally, for signals drawn from
time-translationally invariant distributions. It should be noted
that we also considered boundary conditions other than periodic,
with the aim of illustrating the dependence of boundary conditions
on the extreme statistics. Since time-translational invariance
is violated in these cases, the density of near extreme
states remains open for further studies.

\begin{acknowledgments}
This research has been partly supported by the
Hungarian Academy of Sciences (Grants No.\ OTKA T043734). NRM
gratefully acknowledges support from the EU under a Marie Curie Intra European
Fellowship.
\end{acknowledgments}

\appendix
\section{Derivation of results for $\alpha=4$}
In the case $\alpha=4$, corresponding to random acceleration, the free space and
half space propagators $Z_0$ and $Z_1$ are given by \cite{twb93}
\begin{eqnarray}
Z_0(h,v;h_0,v_0;T)&=&3^{1/2}(2\pi)^{-1}t^{-2}\nonumber\\
&\times&\exp\left\{-3t^{-3}\left[(h-h_0-vt)(h-h_0-v_0t)+{1\over 3}(v-v_0)^2
t^2\right]\right\}\label{Z0alpha=4}\\
Z_1(h,v;h_0,v_0;T)&=&Z_0(h,v;h_0,v_0;T)+\Delta(h,v;h_0,v_0;T)\;,\label{Z1alpha=4}\\
\tilde{\Delta}(h,v;h_0,v_0;s)&=&-{1\over 2\pi}\int_0^\infty{dF\over F^{1/6}}
\int_0^\infty{dG\over G^{1/6}}\nonumber\\
&\times&{\exp\left[-(Fh+Gh_0)-{2\over
3}s^{3/2}\left(F^{-1}+G^{-1}\right)\right]\over F+G}\nonumber\\
&\times&{\rm Ai}\left(-F^{1/3}v+F^{-2/3}s\right){\rm
Ai}\left(G^{1/3}v_0+G^{-2/3}s\right)\;.\label{Delta}
\end{eqnarray}
Here  $\tilde{\Delta}(\dots;s)$ denotes the Laplace transform ${\cal L}\thinspace
\Delta(\dots;T)=\int_0^\infty dT\thinspace e^{-sT}\Delta(\dots;T)$, and ${\rm
Ai}(z)$ is the standard Airy function ${\rm Ai}$ \cite{as}. Using these results, we
rewrite Eq. (\ref{Fperalpha=4}) in the form
\begin{equation}
{\cal F}_{\rm per}(m,T)=1+2\sqrt{\pi}T^{3/2}{\cal L}^{-1}\int_{-\infty}^\infty
dv\thinspace\tilde{\Delta}(m,v;m,v;s)\;,\label{AFper}
\end{equation}
where ${\cal L}^{-1}$ indicates the inverse Laplace transform.

It is convenient to calculate the moments
\begin{equation}
\langle m^\nu\rangle=\int_0^\infty dm\thinspace m^\nu{\partial\over\partial m}{\cal
F}_{\rm per}(m,T)\;,\label{Amoments1}
\end{equation}
which have simple Laplace transforms, and then construct the extreme distribution
from the moments. After substituting Eqs. (\ref{Delta}) and (\ref{AFper}) into
(\ref{Amoments1}), we first integrate over $m$ and then over $v$, using
\begin{eqnarray}
&&\int_{-\infty}^\infty dv\thinspace{\rm Ai}\left(-F^{1/3}v+F^{-2/3}s\right){\rm
Ai}\left(G^{1/3}v+G^{-2/3}s\right)=\nonumber\\
&&\qquad\qquad\qquad(F+G)^{-1/3}{\rm
Ai}\left[s\left(F^{-1}+G^{-1}\right)^{2/3}\right]\;,\label{airyintegral}
\end{eqnarray}
which follows from the integral representation (10.4.32) of ${\rm Ai}(z)$ in Ref.
\cite{as}. On expressing the right-hand side of Eq. (\ref{airyintegral}) in terms of
the Bessel function $K_{1/3}(y)$ according to Eq. (10.4.14) of Ref. \cite{as},
making the changes of variables $G=Fz$, $F={2\over 3}s^{3/2}y^{-1}(1+z^{-1})$,
integrating over $y$ and $z$ with the help of Ref. \cite{gr} and {\it Mathematica},
and using ${\cal L}^{-1}\left[s^{-\alpha-1}\right]=\Gamma(\alpha+1)^{-1}T^\alpha$,
one obtains
\begin{equation}
\langle m^\nu\rangle={\Gamma({3\over 2}\nu+1)\over\Gamma(\nu+1)}\left({T^{3/2}\over
18}\right)^\nu={\Gamma({3\over
2}\nu+1)\over\Gamma(\nu+1)}\left[\Gamma\left(\textstyle{5\over 2}\right)^{-1}\langle
m\rangle\right]^\nu\;,\label{Amoments2}
\end{equation}
which implies the scaled moments (\ref{momentsperalpha=4}).

To calculate $P_{\rm per}(m,T)$, we consider the Laplace transform or generating
function
\begin{equation}
\hat{P}_{\rm per}(\xi,T)=\int_0^\infty dm\thinspace e^{-\xi m}P_{\rm
per}(m,T)=\sum_{n=0}^\infty {\langle m^n\rangle\over n!}(-\xi)^n\;,\label{genfunc}
\end{equation}
Substituting Eq. (\ref{Amoments2}) into (\ref{genfunc}) and carrying out some
straightforward steps, one obtains
\begin{eqnarray}
\hat{P}_{\rm per}(\xi,T)&=&\sum_{n=0}^\infty {\Gamma({3\over
2}n+1)\over\Gamma(n+1)^2} \left[-\Gamma\left(\textstyle{5\over 2}\right)^{-1}\langle
m\rangle
\xi\right]^n\;,\label{step1}\\
&=&\int_0^\infty dx\thinspace e^{-x}\sum_{n=0}^\infty{1\over\Gamma(n+1)^2}
\left[-\Gamma\left(\textstyle{5\over 2}\right)^{-1}\langle m\rangle\xi
x^{3/2}\right]^n\;,\label{step2}\\
&=&\int_0^\infty dx\thinspace e^{-x}J_0\left(2\sqrt{\Gamma\left(\textstyle{5\over
2}\right)^{-1}\langle m\rangle\xi
x^{3/2}}\right)\;,\label{step3}\\
&=&\xi^{-2/3}\int_0^\infty dy\thinspace
\exp\left(-y\xi^{-2/3}\right)J_0\left(2\sqrt{\Gamma\left(\textstyle{5\over
2}\right)^{-1}\langle m\rangle
y^{3/2}}\right)\label{step4}\;,\\
\end{eqnarray}
where $J_0(x)$ is the standard Bessel function \cite{as}.

Next we will need the relation
\begin{equation}
{\cal
L}^{-1}\left[\xi^{-2/3}\exp\left(-y\xi^{-2/3}\right)\right]=m^{-1/3}H(m^{2/3}y)\;,
\label{step5}
\end{equation}
where
\begin{equation}
H(z)=\sum_{n=0}^\infty{(-z)^n\over\Gamma(n+1)\Gamma\left({2\over 3}n+{2\over
3}\right)}\;,\label{step6}
\end{equation}
which follows from expanding the exponential function on the left-hand side of Eq.
(\ref{step5}) in powers of $y\xi^{-2/3}$ and inverting the Laplace transform term by
term. Evaluating the inverse Laplace transform of Eq. (\ref{step4}) using Eq.
(\ref{step5}) and changing the integration variable to $z=m^{2/3}y$ yields
\begin{equation}
P_{\rm per}(m,T)=m^{-1}\int_0^\infty dz\thinspace
H(z)J_0\left(2\sqrt{\Gamma\left(\textstyle{5\over 2}\right)^{-1}m^{-1}\langle
m\rangle z^{3/2}}\right)\;.\label{step7}
\end{equation}

With the help of {\it Mathematica} one may express the series (\ref{step6}) in terms
of generalized hypergeometric functions \cite{gr} and evaluate the integral
(\ref{step7}). This leads to
\begin{eqnarray}
H(z)&=&\Gamma\left(\textstyle{2\over
3}\right)^{-1}\thinspace_0F_4\left(\textstyle{{1\over 3},{1\over 3},{2\over
3},{5\over 6}};-{1\over 108}z^3\right)\\\nonumber &-&\Gamma\left(\textstyle{4\over
3}\right)^{-1}\thinspace z\;_0F_4\left(\textstyle{{2\over 3},{2\over 3},{7\over
6},{4\over 3}};-{1\over 108}z^3\right)\\\nonumber &+&\textstyle{1\over 2}\thinspace
z^2 \;_0F_4\left(1,{4\over 3},{3\over 2},{5\over 3};-{1\over
108}z^3\right)\label{step8}
\end{eqnarray}
and our final result for the extreme distribution,
\begin{equation}
P_{\rm per}(m,T)=\langle m\rangle^{-1}\Phi_{\rm I}
\left({m\over\langle m\rangle}\right)\;,
\end{equation}
where $\Phi_{\rm I}(x)$ is given in Eq. (\ref{Phiperalpha=4}).


\begin{thebibliography}{99}
\bibitem{absorbingboundary} The absorbing boundary condition excludes random
               walks from $Z_1(h_1,h_0,t)$ which reach the boundary $h=0$
           in propagating from $h_0$ to $h_1$.
\end{thebibliography}
\end{document}